# Understanding the Mechanism of the Performance Improvement in Nitrogen-doped Niobium Superconducting Radio Frequency Cavity

Xiaotian Fang,[1] Jin-Su Oh,[1] Matt Kramer,[1] A. Romanenko,[2] A. Grassellino,[2] John Zasadzinski,[3] Lin Zhou[1,4*]

[1.] Division of Materials Science and Engineering, Ames National Laboratory, Ames, IA 50014

[2.] Fermi National Accelerator Laboratory, P.O. Box 500, Batavia, IL, USA

[3.] Department of Physics, Illinois Institute of Technology, IL, USA

[4.] Department of Materials Science and Engineering, Iowa State University, Ames, IA 50014

*Corresponding author: linzhou@ameslab.gov

**Abstract**

Niobium superconducting radiofrequency cavities enable applications in modern accelerators and quantum computers. However, the surface resistance significantly deteriorates the cavities' performance. Nitrogen doping surface treatment can consistently increase cavity performance by reducing surface resistance, but the improvement mechanism is not fully understood. Herein, we employed transmission electron microscopy and spectroscopy to uncover the structural and chemical differences of the Nb/air interface between the non-doped and nitrogen-doped cavities. The results indicate that nitrogen doping passivates the Nb surface by introducing a compressive strain close to the Nb/air interface, which impedes the diffusion of oxygen and hydrogen atoms and reduces surface oxide thickness.

**Impact Statement:** A comprehensive structural and chemical study of the Nb/air interface from the non-doped and nitrogen-doped niobium cavity reveals the performance improvement mechanisms of the nitrogen-doped cavity.

**Introduction**

Niobium superconducting radiofrequency (SRF) cavities are an essential component for particle accelerators [1] and quantum computers [2]. The performance of the SRF cavity is determined by its quality factor (Q-factor) [3], which is inversely proportional to the surface resistance (Rs). Extensive research has been done on the microwave surface resistance of Nb cavities, identifying its critical contribution to radiofrequency (RF) losses at varying frequencies, including the deteriorating Q-factor (Q-drop) with increasing accelerating fields [1,4]. The Rs of the Nb cavities mainly originates in a thin surface layer given by the superconducting Nb penetration depth ~45 nm, with crucial contributions from the Nb/air interface. The complex amorphous oxidation layer [5] and Nb hydrides precipitation on Nb surfaces [6,7] lead to non-superconducting regions that cause serious degradation.

It is of interest to identify all sources of RF losses and to mitigate such problems to improve performance of superconducting RF cavities and qubits. The formation of an amorphous oxidation layer, with a saturated thickness of ~ 6 nm, is unavoidable in the ambient atmosphere without any passivation [8]. The oxidation layer comprises multi-layer (sub-)oxides [3,9,10], where the $Nb_2O_5$ phase is dominant (over ~ 70% of the total thickness) [11]. The suboxides include metallic NbO and semiconducting $NbO_2$, each of which is thermodynamically stable with off-stoichiometry. The mixture of the various oxides at the top surface can increase the surface resistance of the Nb SRF cavity in several ways [12]. NbO is metallic with a superconducting critical temperature $T_C$ ~1.38 K, considerably lower than Nb [13], causing a higher Bardeen–Cooper–Schrieffer (BCS) surface resistance. Vacancies in $Nb_2O_5$ are magnetic [14] and lead to conducting behavior at ~3% concentration, possibly giving rise to nonequilibrium quasiparticle non-two-level-system (TLS) losses [15]. Studies of SRF cavities for 3D qubits reveal evidence of TLS loss and an order of magnitude increase in Q-factor of cavity with the removal of the oxide layer via vacuum annealing [16]. These results indicate that the Nb oxides layer could be a source of both TLS and non-TLS RF losses. Meanwhile, the hydrogen absorption of Nb is usually inevitable since the cavity is processed via multiple polishing procedures during its fabrication [17]. Once hydrogen atoms diffuse into Nb metal, hydride precipitations form near the surface when the metal is cooled below ~170 K [6,18]. These hydrides, with superconducting temperature <1.3 K [7], dramatically decrease the Q-factor of the cavity [19].

Different treatments, such as low-temperature baking [20] and advanced surface polishing strategies [21], have effectively decreased the surface resistance while increasing the Q-factor of SRF cavities in the past few decades. For example, Ciovati [20] baked the cavities between 100 - 200 °C under an ultrahigh vacuum, reducing the Q-drop at high magnetic fields. Copper et al. [22] combined mechanical polishing with light electropolishing on the cavity, which decreased the surface roughness and enhanced the Q-factor to ~$3\times10^{10}$. The rationale is to reduce surface resistance by eliminating the surface oxidation layer. However, a substantial increase in surface resistance has still been observed in varying magnetic fields [20,23], due to the (sub-)oxides and hydrides formed after baking or polishing. Therefore, a method that simultaneously limits the RF losses of the oxidation layer and hydride formation is necessary to further reduce surface resistance.

Nitrogen doping (N-doping) or infusion (N-infusion) has become a predominant method of reducing surface resistance [24,25]. The low-temperature N-infusion, identified by Grassellino et al. [26], exposes the SRF cavities to ~25 mTorr $N_2$ for 48 h at 120 °C, significantly increasing the Q-factor to over ~$5\times10^{10}$, almost twice as the non-infused sample. The surface resistance can be maintained at a minimal value over a range of magnetic fields (0 mT to 80 mT) in N-doped cavities [27]. In addition, significant improvements in the Q-factor of SRF cavities after N-doping/infusion have been observed over a broad range of frequencies (0.65 GHz to 3.9 GHz) [26,28,29]. Thus, the N-doping or N-infusion method has greatly increased the performance of SRF cavities in various applications. However, the mechanism which controls N-doping into Nb cavities and their thermodynamic stability are uncertain. Theoretical simulations [30,31] claim that interstitial N atoms can suppress Nb hydrides' formation and stability, thus improving Q-factor. But there is a lack of experimental validation of the structural and chemical change of the (sub-)surface caused by N-doping/infusion at the atomic scale, the diffusive pathways, and the stability of the N in the near-surface. The unclear mechanistic understanding impedes further development of surface treatment to improve the Q-factor of SRF cavities.

This paper presents a comprehensive structural and chemical investigation of the complex Nb/air interface of the non-doped and N-doped Nb SRF cavities. {100} and {110} Nb facets are observed at the Nb/NbOx interface for both samples. Nitrogen atoms may diffuse into the octahedral interstitial site of the Nb lattice during doping treatment through these facets [32], causing high compressive strain near the Nb surface. The N interstitials and compressive strain impede the diffusion of oxygen and hydrogen atoms, consequently suppressing the oxidation and formation of hydrides. As a result, the oxidation layer thickness of the N-doped sample is ~20% thinner than that of the non-doped sample. The strain relaxation also tends to induce crystallization in the newly formed amorphous NbOx layer. These results clarify the mechanism of the surface resistance reduction by N-doping in the SRF cavity.

**Materials and Methods**

Two groups of Nb cavities, N-doped and non-doped samples, were received from Fermi Lab. The detailed fabrication process of N-doped cavity has been reported in Ref [25, 26]. For local misorientation mapping by electron backscattered diffraction (EBSD), scan was performed under 20 kV and 16 nA electron beam at step size 0.5 μm, with a binning size of 8 × 8. A focused ion beam (FIB, FEI Helios NanoLab G3) was used to prepare the TEM samples with the lift-out technique. A permanent maker was applied as a protective layer before carbon deposition. The microstructure of all the samples was examined using a probe aberration-corrected Titan Themis TEM equipped with a Super-X EDS detector and a quantum 965 GIF system.

**Results and discussion**

The N-doped and non-doped cavities show a clear surface strain difference. Optical microscopy images confirm the grain size of the non-doped sample is ~300 μm to 800 μm, while the grain size of the N-doped sample is ~50 μm to 400 μm (Fig. 1a and c). Interestingly, the backscattered SEM images in Fig. 1b and d show the contrast is uniform within each grain of the non-doped sample, while the contrast varies in the grains of the N-doped sample. This irregular contrast suggests the existence of grain misorientation or a change in the degree of deformation near the surface in the N-doped cavities [33]. To quantitatively analyze the local misorientation, the EBSD result is obtained from a ~250 μm × 250 μm region within one grain both from non-doped and N-doped samples, as shown in Fig. 2a and d. The inverse pole figure (IPF) coloring map shows only one color for the region in the non-doped sample (Fig. 2b). In contrast, several domains with different colors are discovered in the N-doped sample (Fig. 2e), indicating the formation of subgrains or low-angle boundaries. The kernel average misorientation (KAM) mapping in Fig 2c and f, representing the local misorientation, is often used for strain analysis [34]. The local misorientation of the N-doped sample is significantly higher than that of the non-doped sample, indicating the higher strain level at the surface by N-doping [34], which is consistent with the SEM results. The N-doped sample's high surface strain is likely due to the diffusion of N atoms into the Nb lattice during surface treatment [22], similar to the hardening process for carburized steel with high in-plane compressive strain near the surface [35].

A relatively uniform ~ 6 nm amorphous NbOx layer is revealed in the non-doped sample (Fig. 3a). Corresponding energy-dispersive X-ray spectroscopy (EDS) elemental mappings (Fig. 3b-d) display uniform O and Nb distribution inside the NbOx layer. The Nb metal shows {110} or {100} facets at the Nb/NbOx interface by atomic resolution bright-field (BF) STEM image, as indicated in Fig. 3e. These facets are formed in the close-packed planes in Nb to minimize interface energy [36]. Both facets have a 2~4 monolayers step at the interface and the steps generate a 4~6 monolayers terrace. The morphology of the facets (steps and terraces) tends to have different superconducting gaps, thus impacting the cavities' performance [33]. Meanwhile, the {110} and {100} facets provide a larger interface area for the diffusion [37,38]. Since O or H atoms prefer to diffuse into the interstitial sites of the Nb lattice (schematic in Fig. 3b) [30,39,40], the facets at the Nb/NbOx interface could facilitate subsequent formation of Nb (sub-)oxides or hydrides.

Some interstitial oxygen atoms are visible by angular bright-field (ABF) imaging, which is an imaging mode that is sensitive to light elements by using an annular detector with a collection angle of 11 mrad-22 mrad [41]. Figure 3g shows an atomic resolution ABF image around the NbOx/Nb interface. The O atomic positions are labeled by blue arrows, consistent with the schematic in Fig. 3f. The interstitial O atoms are stable and may not be released below 1600 °C [10]. Moreover, these interstitial O atoms distort the Nb lattice and introduce strain at the interface, as demonstrated by the lattice image in Fig. 3e.

There is a transition of the valence state of Nb in the NbOx layer across the NbOx/Nb interface, as shown by the electron energy loss spectra (EELS) in Fig. 3h-j. The Nb-$M_{2,3}$ and O-K edge can distinguish different oxides (NbO, $NbO_2$, or $Nb_2O_5$) [12]. The Nb $M_2$ and $M_3$ edge is mainly attributed to the transition of Nb $3p$ electrons to unoccupied Nb $4d$ and $5s$ states, where the spin-orbit coupling of the $3p$ electrons results in the two sharp edges, namely the $M_3$ and $M_2$ [12]. With the valence changes from Nb (0) to Nb (+5), a right shift of the $M_3$ or $M_2$ edges will be observed. An EELS line scan across the NbOx (step size ~ 0.3 nm) and the color map generated by the first derivative method of the spectrum [42] are shown in Fig. 3h. There is an apparent shift of Nb-$M_3$ and O-K edge position across the Nb/$NbO_x$ interface. After moving ~1.5 nm from the interface into the NbOx layer, the Nb-$M_3$ and O-K edges shift again. These shifts imply the valance changes at the two positions, indicating the thickness of the suboxides-rich layer is ~1.5 nm. A detailed point analysis is subsequently applied to similar regions, and the results are shown in Fig. 3i-j. The shift of the $M_3$ edge and gradual change of the shoulders of Nb-$M_3$ (arrows in Fig. 3i) indicate an increasing phase fraction of $Nb_2O_5$ [43]. As for the O-K edge, the double-peak feature near the edge onset becomes visible from points 1 to 4, resulting from the hybridization of oxygen $2p$ and Nb $4d$ orbitals, and indicates the formation of $Nb_2O_5$ at position 4. Meanwhile, the gradually increasing intensity of the peaks near ~545 eV suggests an increase in the valence of Nb [43]. Therefore, the EELS line scan and point analysis validate a suboxide layer at the Nb/amorphous NbOx interface.

The above results indicate that the oxide layer's formation on the Nb surface is diffusion-controlled: the O atoms continuously diffuse into Nb octahedral interstitials until reaching saturated thickness [8]. And sub-oxides will be formed due to the deficiency of O atoms near the (sub-)surface.

The NbOx layer thickness of the N-doped sample is ~4.8 nm (Fig. 4a). Corresponding EDS elemental mappings (Fig. 4b-d) also show intermixing O and Nb signals inside the NbOx layer. No clear nitrogen signal is detected in the Nb metal, due to the low concentration of $N_2$ during surface treatment [31]. The {110} and {100} facets are discovered at the Nb/NbOx interface (Fig. 4e), showing the same critical planes for N (O /H) atoms diffusion as the non-doped sample. The transition of the valence state of Nb in the NbOx layer also exists in the N-doped sample, as shown in Fig. 4f-h. The EELS line profile in Fig. 4f shows a shift at the Nb/NbOx interface, and another shift after moving ~1 nm from the interface into the NbOx layer, indicating the thickness of the suboxide-rich region is ~ 1nm in the N-doped sample. Point analysis, as shown in Fig. 4g-h, confirms an apparent shift of Nb $M_3$ and O-K peak, suggesting the phase fraction of $Nb_2O_5$ gradually increases from Nb to NbOx surface.

Although the NbOx layer shares chemical similarity in N-doped and non-doped samples, crystallized features are occasionally observed in the N-doped sample, as shown in Fig. 5. A ~5 nm nanocrystallized domain is captured in the N-doped sample. The spacing of the lattice fringes is ~0.24 nm, close to the d-spacing of NbO {111} planes (0.2425 nm). EELS analysis also identified that the nanocrystal is highly possible to be NbO. The crystallized feature is not observed in the NbOx layer of the non-doped sample. Meanwhile, previous research indicates that the whole NbOx layer is removed during the nitrogen-doping treatment [25]. Thus, the crystallized domains are formed during post-treatment in ambient conditions. This crystallization of NbO may be induced by the Nb surface strain relaxation [44–46]. The compressive strain close to the Nb surface may also increase the hardness of the surface oxide layer [47,48].

After comparing the microstructural and chemistry of non-doped and N-doped samples, we attribute the improved property of the N-doped sample to the following reasons. First, the thickness of the oxidation layer of the N-doped sample is $4.7 \pm 0.6$ nm, ~20% less than that of the non-doped sample ($5.8 \pm 0.6$ nm), resulting in lower surface resistance in the N-doped sample. The thickness reduction of the NbOx layer is related to the diffusion-control oxidation procedure. The Nb-oxide layer is removed at a high temperature before N-doping [25]. N atoms can diffuse into the octahedral interstitials of the Nb, causing a large residual in-plane compressive strain near the surface. The N-occupied interstitials and compressive strain impede the diffusion of O atoms after treatment, subsequently reducing the thickness of the NbOx layer. Meanwhile, strain relaxation induces a small amount of crystallization in the newly formed NbOx layer. The nano-domains potentially block the movement of O atoms, limiting the mobility of free O atoms in the NbOx layer. Finally, the H atoms diffusion will also impede by a similar mechanism, hindering the formation of hydrides. The interstitial N atoms decrease hydrides' thermodynamic stability [30], reducing the likelihood of hydride precipitation at low temperatures. As a result, the cavities' surface resistance decreases, and the Q-factor increases in various magnetic fields and frequencies by the N-doped surface treatment.

**Conclusion**

Our results reveal the mechanisms of the property improvement in cavities by the $N_2$ doping surface treatment. On the one hand, the N atoms diffuse into the Nb lattices, producing compressive strain near the surface. The interstitial N atoms and compressive strain impede the diffusion of oxygen atoms. We observed ~20% thickness reduction of the oxidation layer in the N-doped sample compared with the non-doped sample. On the other hand, the interstitial N atoms could reduce the diffusion of hydrogen atoms and the stability of the hydrides at low temperatures. Since the amorphous oxidation layer and hydrides are two major components of the surface resistance, the N-doping surface treatment efficiently increases the Nb cavity's performance. These findings will facilitate surface treatment optimization of the Nb SRF cavity.

**Acknowledgments**

This work was supported by the U.S. Department of Energy, Office of Science, National Quantum Information Science Research Centers, Superconducting Quantum Materials and Systems Center (SQMS) under the contract No. DE-AC02-07CH11359. All electron microscopy and related work were performed using instruments in the Sensitive Instrument Facility in Ames Lab. The Ames

Laboratory is operated for the U.S. Department of Energy by Iowa State University under Contract No. DE-AC02-07CH11358.13

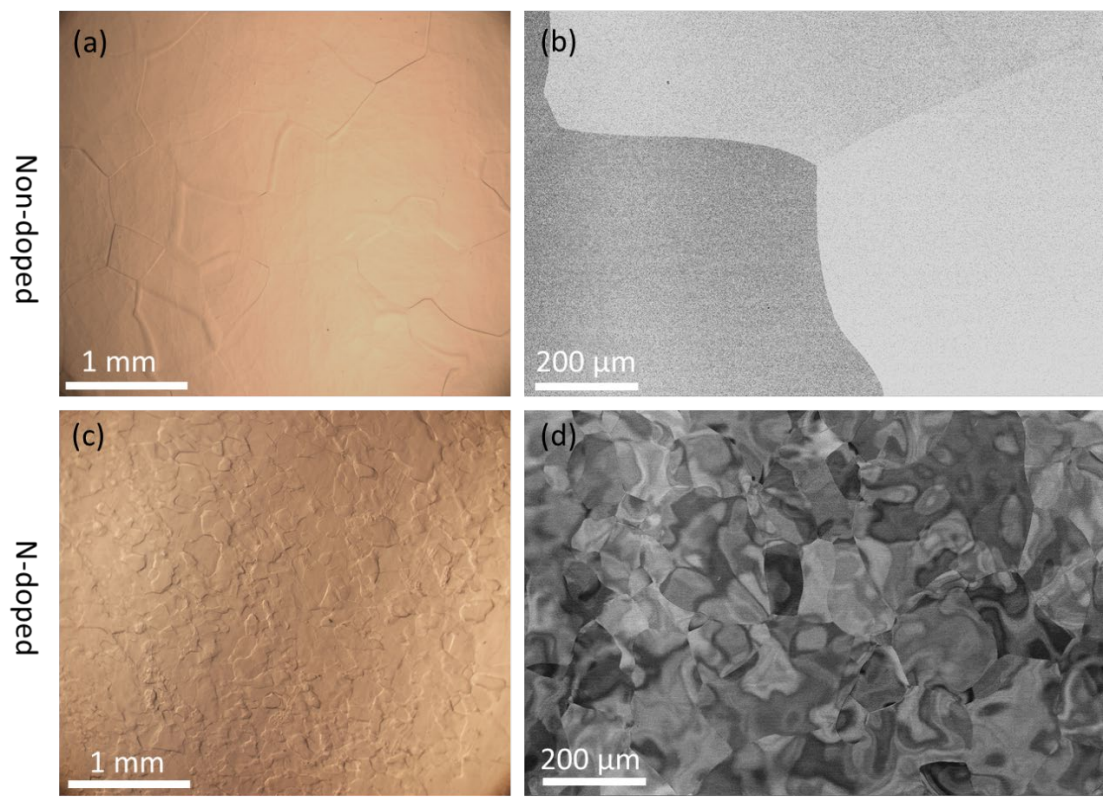

**Figure 1. Surface morphology characterization.** The optical (a, c) and back scattered SEM (b, d) images of the non-doped (a-b) and nitrogen-doped (c-d) samples.

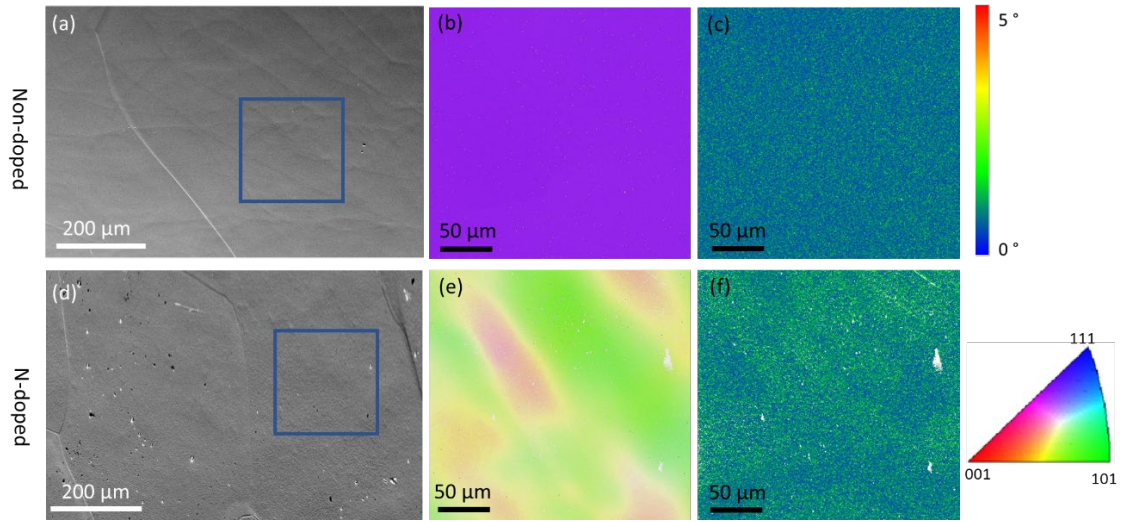

**Figure 2. EBSD analysis of local misorientation.** (a-c) non-doped sample and (d-f) N-doped sample. The blue squares in (a) and (d) indicate the ~ 250 μm × 250 μm region within a coarse grain for EBSD scanning; IPF map of (b) non-doped and (e) N-doped samples, the domains with different colors in (e) represent regions with different orientations; (c) and (f) the corresponding KAM mapping of (b) and (e).

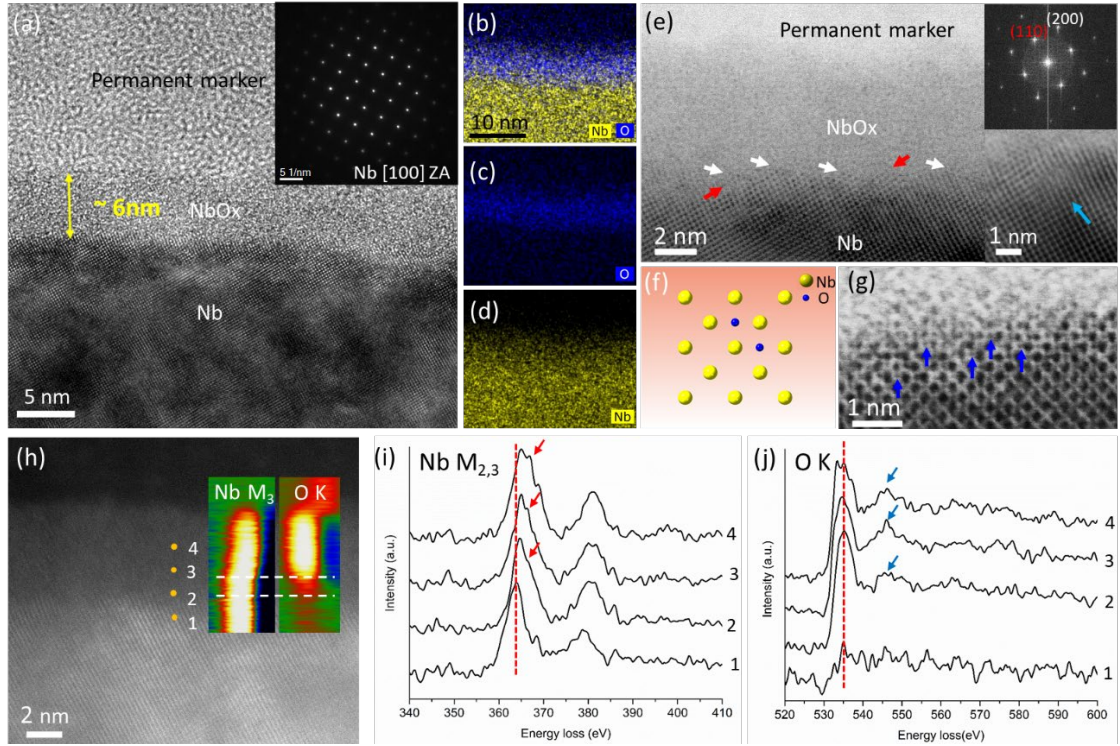

**Figure 3. Microstructural and chemical analysis at Nb surface of the non-doped sample.** (a) HRTEM image taken along [100] zone axis (ZA) and (b-d) EDS elemental mapping of Nb/air interface. (e) High-resolution BF-STEM image (top right inset is the corresponding fast Fourier transformation (FFT) image) shows the NbOx/Nb interface area. The white and red arrows indicate the {100} and {110} facets. The inset at the bottom right shows Nb lattice distortion close to the NbOx/Nb interface (indicated by light blue arrow). (f) schematic of Nb crystal structure with interstitial oxygen at the octahedral center, viewed along [100] ZA; (g) ABF image near the NbOx/Nb interface. The oxygen interstitials are indicated by blue arrows. (h-j) EELS analysis of the NbOx layer: (h) HAADF-STEM image and the first derivative of EELS line scan (color inset) of the Nb-$M_3$ and O-K edges; (i-j) EELS spectrum of Nb-$M_{2,3}$ and O-K edges from positions 1 to 4 in (h).

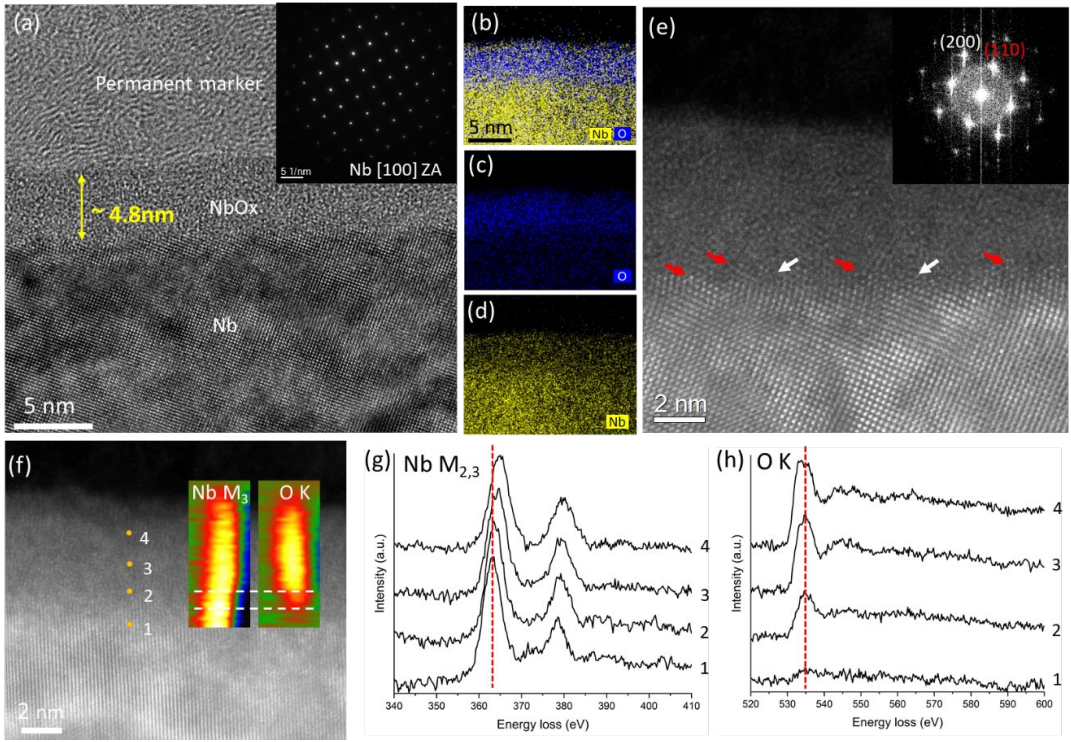

**Figure 4. Microstructural and chemical analysis at Nb surface of N-doped sample.** (a) HRTEM image taken along [100] ZA and (b-d) EDS mapping of Nb/air interface. (e) High-resolution HAADF-STEM image shows the NbOx/Nb interface area (top right inset is the corresponding FFT), red and white arrows indicate the {110} and {100} facets. (f-h) EELS analysis of the NbOx layer: (f) HAADF-STEM image and the first derivative of EELS line scan (color inset) of the Nb-$M_3$ and O-K edges; (g-h) EELS spectrum of Nb-$M_{2,3}$ and O-K edges from position 1 to 4 in (f).

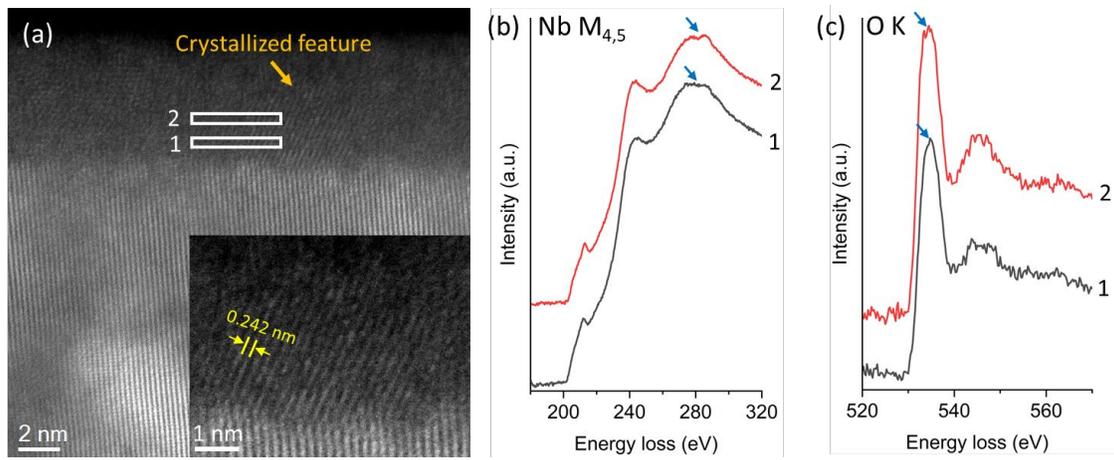

**Figure 5. Crystallized feature in NbOx amorphous layer.** (a) HRSTEM image shows a crystalized feature in the amorphous layer, amplifying in the inset. (b-c) the EELS data from regions 1 and 2.